\documentclass[aps,pre,twocolumn,showpacs,longbibliography]{revtex4-1}
\usepackage{bbding}
\usepackage{amsmath}
\usepackage{amsthm}
\usepackage{graphicx}
\usepackage{subfigure}
\usepackage{amssymb}
\usepackage{graphicx}
\usepackage{dcolumn}
\usepackage{bm}
\usepackage{float}
\usepackage[colorlinks,linkcolor=blue,citecolor=blue,urlcolor=blue,hyperindex,breaklinks]{hyperref}

\begin{document}

\title{Multifunctional quantum thermal device utilizing three qubits}
\author{Bao-qing Guo}
\author{Tong Liu}
\author{Chang-shui Yu}
\email{ycs@dlut.edu.cn}
 \affiliation{School of Physics, Dalian University of Technology, Dalian 116024, China}%
\date{\today }

\begin{abstract}
Quantum thermal devices which can manage heat as their electronic analogues
for the electronic currents have attracted increasing attention. Here a
three-terminal quantum thermal device is designed by three coupling qubits
interacting with three heat baths with different temperatures. Based on the
steady-state behavior solved from the dynamics of this system, it is
demonstrated that such a device integrates multiple interesting
thermodynamic functions. It can serve as a heat current \textit{transistor}
to use the weak heat current at one terminal to effectively amplify the
currents through the other two terminals, to continuously 
modulate  them ranging in a large amplitude, and even to switch on/off the heat currents. It is
also found that the three currents are not sensitive to the fluctuation of the temperature
at the low temperature terminal, so it can behave as a thermal \textit{stabilizer}. In addition, we can utilize
one terminal temperature to ideally turn off the heat current at any one terminal and to allow the heat currents through the other two terminals, so it can be used as a thermal \textit{valve}. Finally, we illustrate that this thermal device can control the heat 
currents to flow unidirectionally, so it has the function as a thermal \textit{rectifier}.
\end{abstract}

\maketitle

\affiliation{School of Physics, Dalian University of Technology, Dalian
116024, China}

\affiliation{School of Physics and Optoelectronic Technology, Dalian University of
Technology, Dalian 116024, China}

\section{Introduction}

Quantum thermodynamics, as the combination of the classical thermodynamics
and the quantum theory, has attracted increasing interest in recent years.
Various quantum engines and quantum refrigerators as well as some particular
quantum thermodynamical devices have been studied extensively. These provide
not only the fundamental physical platforms to test the macroscopic
thermodynamic laws down to the quantum level, but also give valuable
references to design the microscopic quantum devices with some particular
functions which could be used to purposively manage the heat currents.

As we know, an electronic diode \cite{Lashkaryov1941} consisting of two
terminals guide current to unidirectionally flow, while a transistor \cite%
{Bardeen1998} owing three terminals can control currents through two
terminals by manipulating the third terminal current so that to realize
three basic functions: a switch, an amplifier, or a modulator. They, used to
effectively manage the electricity or for logical operations, have led to
the electronic information revolution since the last century. How to control
the thermal transport is also a key challenge of the modern technology in
energy conversion systems such as heating and refrigeration, thermal
management and so on. For example, quantum heat engines and refrigerators
have been investigated theoretically and experimentally for a long time to
study their efficiencies and to test the laws of thermodynamics \cite%
{Levy2012,Feldmann2000,Palao2001,Arnaud2002,Segal2006,DeTomas2012,Geva1992,
Geva1996,Kosloff2010,Thomas2011,Feldmann1996,Feldmann2003,Quan2007,Linden2010,He2017,Yu2014,Man2017,Silva2015,Abah2012,Rosnagel2016,Alicki1979,Skrzypczyk2011,Qin2017}%
. Recently, some thermal diodes and transistors, analogous to their
electronic counterparts, have been designed based on various phase change
materials such as VO2 \cite%
{Ito2016,Yang2013,Ito2014,Ben-Abdallah2014,Joulain2015,Wehmeyer2017}. In
particular, many quantum mechanical thermal diodes and transistors have also
been proposed in terms of different systems \cite%
{Marcos-Vicioso2018,Maznev2013,Werlang2014,Chen2015,Li2004,Pereira2011,Wang2012,Kobayashi2009,Fratini2016,Landi2014,Man2016,Jiang2015,Guo2018,Joulain2016,Lo2008,Li2006,Komatsu2011,Segal200501}. In addition, some quantum devices including thermal valve \cite{Zhong2012},
logic gates \cite{Wang2007} and memory \cite{Wang2008} have also been
reported for a potential way to quantum information processing, and some
other devices like quantum thermal ratchet \cite{Faucheux1995,Zhan2009},
stabilizer \cite{Guo2018}, thermometer \cite{Hofer2017} and batteries \cite%
{Binder2015,Campaioli2017,Ferraro2018} have also been presented, which
further enriches the potential applications of quantum thermodynamic
systems. However, one can easily find that most of the previously proposed
quantum thermal devices usually realize a relatively unique function. So how
we can realize multiple functions by a single quantum dynamics mechanics is
what we are interested in in this paper.

Motivated by this quest, we design a multifunctional quantum thermal device
by utilizing the strong internal coupling three qubits. Every qubit in our
system is connected to a heat bath with a given temperature. We apply the perturbative secular master equation for open system to study the steady-state thermal behaviors in detail\cite{Breuer2002}. It is shown that our system can serve as a 
\textit{transistor}, that is, a weak heat current at one terminal can
significantly \textit{amplify} the heat currents through the other two
terminals ones. At the same time, with the weak heat current changed
slightly, the heat currents through the other two thermals can also be 
\textit{modulated} continuously ranging from a small value to a large one.
In particular, if one heat current is weak enough, the heat currents
through the other two thermals can be well restricted below a small
threshold value (i.e., ``cut off"), which acts as a \textit{switch}. In
addition, we show that the heat currents are very robust to the temperature
fluctuation at the lowest temperature terminal, so this system can be used as
a thermal \textit{stablizer}. It is quite interesting that our system can
also act as a good thermal \textit{valve} which can perfectly cut off the
heat current at any one terminal and allow the heat to flow through the
other two terminals. Finally, we demonstrate that our system can also be
used to \textit{rectify} the heat current when we block the heat current at
one terminal. The remaining of the paper is organized as follows. In Sec. %
\ref{sec:modelandME}, we present the model of our system, and give the
dynamics by applying the master equation and then solve the steady state. In
Sec. \ref{sec:multiplefunctions}, we demonstrate the various thermodynamical
functions by analyzing the thermal behaviors in the steady-state case. We
give a discussion about a possible experimental realization and the other possible energy level configurations, and finally conclude our work in Sec. \ref{sec:conclusion}.

\section{\label{sec:modelandME} The model and the dynamics}

\begin{figure}[tbp]
\centering
\includegraphics[width=0.75\columnwidth]{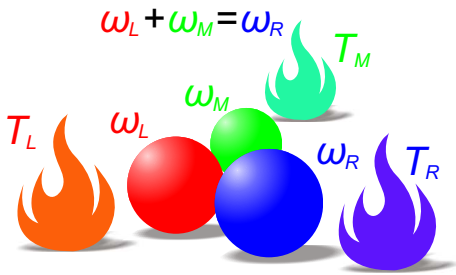}
\caption{(Colour online) Three coupling qubits with the transition
frequencies $\protect\omega_L$, $\protect\omega_M$, and $\protect\omega_R$
are in contact with three separate baths at temperatures $T_L$, $T_M$, and $%
T_R$, where $\protect\omega_L + \protect\omega_M = \protect\omega_R$ means
the resonant coupling.}
\label{fig:modeldiagram}
\end{figure}

Let us consider that three coupling qubits interact with three heat baths,
as is sketched in Fig.~\ref{fig:modeldiagram}. The transition frequencies of
the three qubits are denoted by $\omega _{L}$, $\omega _{M}$ and $\omega
_{R} $, and the temperatures of the three baths are represented by $T_{L}$, $%
T_{M} $, and $T_{R}$, where the subscripts correspond to the qubits they are
in contact with. Here we suppose that three qubits resonantly interact with
each other, that is, $\omega _{L}+\omega _{M}=\omega _{R}$ is implied and
without loss of generality, we let $\omega _{R}>\omega _{L}>\omega _{M}$. In
such a model, the only resources driving the model to work are the three
heat baths. One will see that such a model will act as a multifunctional
quantum thermal device as considered throughout the paper. To show this, we
will have to begin with the dynamics of the open system.

The Hamiltonian of the three interacting qubits reads 
\begin{equation}
H_{S} = H_{0} + H_{I} ,
\end{equation}
where the free Hamiltonian is 
\begin{equation}
H_{0} = \sum_{\mu=L,M,R} \frac{1}{2}\omega_{\mu}\sigma_{\mu}^{z},
\end{equation}
and the resonant internal interaction Hamiltonian is 
\begin{equation}
H_{I} = g \sigma_{L}^{x}\sigma_{M}^{x}\sigma_{R}^{x},
\end{equation}
with $\sigma^{x}$ and $\sigma^{z}$ denoting the Pauli matrices, $g$ denoting
the coupling strength. Hereinafter we set the Planck constant and the
Boltzmann constant to be unit, i.e., $\hbar=k_{B}=1$ for simplicity. Note
that this type interaction Hamiltonian has been proposed and studied in
some spin systems \cite{Reiss2003,Pachos2004,Bermudez2009,Seah2018}. Furthermore
let the three qubits contact with three heat baths with the Gibbs state $%
\rho_{\mu}=\exp{(-H_{\mu}/T_{\mu})}/{\mathrm{Tr}[\exp{(-H_{\mu}/T_{\mu})}]}$
where $H_{\mu}=\sum_{k}\omega_{\mu k} b^{\dagger}_{\mu k} b_{\mu k}$, $\mu =
L, M, R$ respectively, $\omega _{\mu k}$ and $b_{\mu k}$ denote the
frequencies and the annihilation operators of the bath mode with $[b_{\mu
k}, b_{\nu k^{^{\prime }}}^{\dagger }]=\delta _{\mu ,\nu }\delta
_{k,k^{^{\prime }}}, [b_{\mu k}^{\dagger }, b_{\nu k^{^{\prime }}}^{\dagger
}]=0, [b_{\mu k}, b_{\nu k^{^{\prime }}}]=0$. The interaction Hamiltonian
between the system and the baths is given by 
\begin{equation}
H_{SB}=\sum_{\mu k} f_{\mu k}\sigma_{\mu}^{x}(b_{\mu k} + b_{\mu
k}^{\dagger}),
\end{equation}
where $f_{\mu k}$ stands for the coupling strength between the $\mu$th qubit
and the $k$th mode in its bath. Thus the total Hamiltonian of the whole
system including the baths and the qubits can be written as 
\begin{equation}  \label{Htotal}
H_{total}= H_{S} + \sum_{\mu} H_{\mu} + H_{SB}.
\end{equation}

To derive the master equation that governs the evolution of our open system,
we have to turn to the $H_{S}$ representation. To do so, let's consider the
eigen-decomposition of $H_{S}$ as 
\begin{equation}
H_{S}=\sum_{k}\lambda _{k}\left\vert \lambda _{k}\right\rangle \left\langle
\lambda _{k}\right\vert 
\end{equation}%
where the eigenvalues read 
\begin{equation}
-\lambda _{1+j}=\lambda _{8-j}=\sqrt{\Lambda _{1+j}^{2}+g^{2}},j=0,1,2,3,
\end{equation}%
with $[\mathit{\Lambda }_{1},\mathit{\Lambda }_{2},\mathit{\Lambda }_{3},%
\mathit{\Lambda }_{4}]=[\omega _{R},\omega _{L},\omega _{M},0]$, and the
eigenvectors are explicitly given in the Appendix. Thus the interaction
Hamiltonian $H_{SB}$ in $H_{S}$ representation can be rewritten as 
\begin{equation*}
H_{SB}=\sum_{\mu ,k,l}f_{\mu k}(V_{\mu l}(\omega _{\mu l})+V_{\mu
l}^{\dagger }(\omega _{\mu l}))(b_{\mu k}+b_{\mu k}^{\dagger }),
\end{equation*}%
where the eigenoperator $V_{\mu l}(\omega _{\mu l})$ of $H_{S}$ and their
corresponding to the eigenfrequencies $\omega _{\mu l}$ satisfy the relation 
$[H_{S},V_{\mu l}(\omega _{\mu l})]=-\omega _{\mu l}V_{\mu l}(\omega _{\mu
l})$ and their explicit expressions are also given in the Appendix.
Therefore, following the standard procedure \cite{Breuer2002},  one can
apply the Born-Markovian approximation and the secular approximation to
obtain the master equation \cite{Breuer2002} as 
\begin{equation}
\dot{\rho}=-\mathrm{i}[H_{S},\rho ]+\mathcal{L}_{L}[\rho ]+\mathcal{L}%
_{M}[\rho ]+\mathcal{L}_{R}[\rho ],  \label{eq:masterequation}
\end{equation}%
where $\rho $ is the density matrix of the system and the Lindblad operator $%
\mathcal{L}_{\mu }[\rho ]$ is given by 
\begin{align}
\mathcal{L}_{\mu }[\rho ]& =\sum_{l}J_{\mu }(-\omega _{\mu l})[2V_{\mu
l}(\omega _{\mu l})\rho V_{\mu l}^{\dagger }(\omega _{\mu l})  \notag \\
& -\{V_{\mu l}^{\dagger }(\omega _{\mu l})V_{\mu l}(\omega _{\mu l}),\rho \}]
\notag \\
& +J_{\mu }(+\omega _{\mu l})[2V_{\mu l}^{\dagger }(\omega _{\mu l})\rho
V_{\mu l}(\omega _{\mu l})  \notag \\
& -\{V_{\mu l}(\omega _{\mu l})V_{\mu l}^{\dagger }(\omega _{\mu l}),\rho
\}],
\end{align}%
with the spectral densities defined by 
\begin{equation}
J_{\mu }(\pm \omega _{\mu l})=\pm \gamma _{\mu }(\omega _{\mu l})n(\pm \omega
_{\mu l}),  \label{eq:JP}
\end{equation}%
and the average thermal excitation number defined by 
\begin{equation}
n(\omega _{\mu l})=\frac{1}{\mathrm{e}^{\frac{\omega _{\mu l}}{T_{\mu }}}-1}
\label{eq:photonnumber}
\end{equation}%
subject to the frequency $\omega _{\mu l}$ and the temperature $T_{\mu }$.
During the derivation of the master equation, the secular approximation
requires the relaxation time of the system $\tau _{R}\sim 1/\gamma _{\mu
}(\omega _{\mu l})$ is large compared to the typical time scale of the
intrinsic evolution of the system $\tau _{S}\sim {|\omega _{\mu l}-\omega
_{\mu l^{\prime }}|}^{-1}$. So we have the condition $\gamma _{\mu }(\omega
_{\mu l})\ll \{|\omega _{\mu l}-\omega _{\mu l^{\prime }}|\}$ which
signifies that the strong internal coupling strength greatly separates the energy levels. This is consistent with the conclusion in references \cite{Hofer2017NJP,Gonzalez2017,Rivas2010,Seah2018} where the valid internal coupling strength regime is discussed via different master equations, such as local, global and coarse-graining master equations. Most importantly, the global master equation in the strong internal coupling strength regime coincides well with the laws of thermodynamics as shown in the above references.  Definitely different bath spectra lead to different physical phenomenons \cite{Breuer2002,Valleau2012}. In the following text we assume $\gamma _{\mu}(\omega_{\mu l})=\gamma _{\mu }$ does not depend on the transition frequency
for simplicity. Note that we also have tested the Ohmic bath spectrum and found that similar quantum thermal functions can be achieved given appropriate parameters. Only the difference between the valves using the two different bath spectra is present in Fig.~\ref{fig:valve}.

The dynamical behavior of the system at any time is determined by the master
equation Eq.~(\ref{eq:masterequation}). However, we concern its behaviour
at the steady state in order to construct thermal device about heat current.
Therefore what to do first is to solve the steady state solution of Eq.~(\ref%
{eq:masterequation}), i.e. $\dot{\rho}_{S}=0$. After some arrangement of
steady state solution, one can obtain a system of linear equations about the
elements of the density matrix as 
\begin{equation}
\sum_{\mu =M,L,R}\mathbf{M}_{\mu }\left\vert \rho \right\rangle =0,\quad
\rho _{ij}^{S}=0,i\neq j,\label{eq:rhodiagonal}
\end{equation}%
where $\left\vert \rho \right\rangle ^{T}=[\rho _{11}^{S},\rho
_{22}^{S},...,\rho _{88}^{S}]$ with {\allowdisplaybreaks[4] 
\begin{align}
\mathbf{M}_{L}& =(C_{1,1;3,2}+C_{6,1;8,2})\mathbf{J}_{L1}(C_{1,1;3,2}^{%
\dagger }+C_{6,1;8,2}^{\dagger })  \notag \\
& +(C_{1,1;6,2}+C_{3,1;8,2})\mathbf{J}_{L2}(C_{1,1;6,2}^{\dagger
}+C_{3,1;8,2}^{\dagger })  \notag \\
& +(C_{2,1;4,2}+C_{5,1;7,2})\mathbf{J}_{L3}(C_{2,1;4,2}^{\dagger
}+C_{5,1;7,2}^{\dagger })  \notag \\
& +(C_{2,1;5,2}+C_{4,1;7,2})\mathbf{J}_{L4}(C_{2,1;5,2}^{\dagger
}+C_{4,1;7,2}^{\dagger }), \\
\mathbf{M}_{M}& =(C_{1,1;2,2}+C_{7,1;8,2})\mathbf{J}_{M1}(C_{1,1;2,2}^{%
\dagger }+C_{7,1;8,2}^{\dagger })  \notag \\
& +(C_{1,1;7,2}+C_{2,1;8,2})\mathbf{J}_{M2}(C_{1,1;7,2}^{\dagger
}+C_{2,1;8,2}^{\dagger })  \notag \\
& +(C_{3,1;4,2}+C_{5,1;6,2})\mathbf{J}_{M3}(C_{3,1;4,2}^{\dagger
}+C_{5,1;6,2}^{\dagger })  \notag \\
& +(C_{3,1;5,2}+C_{4,1;6,2})\mathbf{J}_{M4}(C_{3,1;5,2}^{\dagger
}+C_{4,1;6,2}^{\dagger }), \\
\mathbf{M}_{R}& =(C_{1,1;4,2}+C_{5,1;8,2})\mathbf{J}_{R1}(C_{1,1;4,2}^{%
\dagger }+C_{5,1;8,2}^{\dagger })  \notag \\
& +(C_{1,1;5,2}+C_{4,1;8,2})\mathbf{J}_{R2}(C_{1,1;5,2}^{\dagger
}+C_{4,1;8,2}^{\dagger })  \notag \\
& +(C_{2,1;3,2}+C_{6,1;7,2})\mathbf{J}_{R3}(C_{2,1;3,2}^{\dagger
}+C_{6,1;7,2}^{\dagger })  \notag \\
& +(C_{2,1;6,2}+C_{3,1;7,2})\mathbf{J}_{R4}(C_{2,1;6,2}^{\dagger
}+C_{3,1;7,2}^{\dagger }).
\end{align}%
}Here $\mathbf{J}_{\mu l}=
\begin{pmatrix}
1 & 0 \\ 
0 & 0%
\end{pmatrix}\otimes 
\begin{pmatrix}
1 & 0 \\ 
0 & 0%
\end{pmatrix}\otimes 
\begin{pmatrix}
-B_{\mu l} & A_{\mu l} \\ 
B_{\mu l} & -A_{\mu l}%
\end{pmatrix}%
$ with $A_{\mu l}=\exp (\omega _{\mu l}/T_{\mu })B_{\mu l}$, $B_{\mu
l}=\gamma _{\mu }n(\omega _{\mu l}){\sin }^{2}{\alpha _{\mu l}}$, and $%
C_{i,j;m,n}=\left\vert i\right\rangle \left\langle j\right\vert +\left\vert
m\right\rangle \left\langle n\right\vert $ with $\left\{ \left\vert
i\right\rangle \right\} $ representing the natural orthonormal basis of $8$%
-dimensional Hilbert space. The negative sign in $\mathbf{J}_{\mu l}$
denotes the population decrement from a relevant level while the positive
sign means the population increment. Based on the steady state solution, one
can obtain the heat currents as \cite{Levy2012,Szczygielski2013,Kolar2012} 
\begin{equation}
\dot{Q}_{\mu }=\mathrm{Tr}(H_{S}\mathcal{L}_{\mu }[\rho ^{S}])=\left\langle
\lambda \right\vert \mathbf{M}_{\mu }\left\vert \rho \right\rangle ,
\label{eq:currentdefinition}
\end{equation}%
originating from the dissipation of the $\mu $th bath, where ${\left\vert
\lambda \right\rangle }^{T}=[\lambda _{1},\lambda _{2},\lambda
_{3},...,\lambda _{8}]$. One should note that $\dot{Q}_{\mu }>0$ denotes the
system absorbing heat from the $\mu $th bath, while $\dot{Q}_{\mu }<0$ means
that the heat flows into the bath. So the remaining key task is to solve the
steady state solution of the master equation Eq.~(\ref{eq:rhodiagonal}).
However, the analytical solution of Eq.~(\ref{eq:rhodiagonal}) is so tedious
that we cannot explicitly give it here, so we would like to demonstrate the
various thermodynamic functions based on the numerical solution in the next
section.

\section{The various thermodynamic functions}

\label{sec:multiplefunctions}

The essence of a quantum thermal device is that the heat currents can be purposively controlled. In this section, we will show
that our system  can work as
a thermal device with multiple different thermodynamic functions such as \textit{amplifier}, \textit{modulator}, \textit{switcher}, \textit{%
valve}, \textit{stabilizer}, and \textit{rectifier}.
\begin{figure}[tbp]
\centering
\includegraphics[width=0.9\columnwidth]{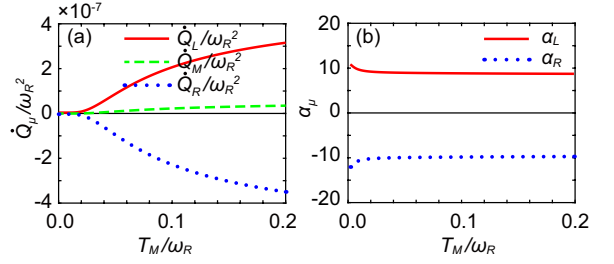}
\caption{(Colour online) (a) Three heat currents $\dot{Q}_{\protect\mu}/%
\protect\omega_{R}^2$ versus $T_M/\protect\omega_{R}$ at the steady state.
The red solid, green dashed, and blue dotted lines correspond to the heat currents $%
\dot{Q}_{L}/\protect\omega_{R}^2$, $\dot{Q}_{M}/\protect\omega_{R}^2$, and $%
\dot{Q}_{R}/\protect\omega_{R}^2$, respectively. (b) The amplification factors $%
\protect\alpha_{\protect\mu}$ versus $T_{M}/\protect\omega_{R}$ at the
steady state. The red solid and blue dotted lines correspond to $\protect%
\alpha_{L}$ and $\protect\alpha_{R}$, respectively. Here $\protect\omega%
_{L}=0.9\protect\omega_{R}$, $\protect\omega_{M}=0.1\protect\omega_{R}$, $%
g=0.1\protect\omega_{M}$, $\protect\gamma_{L}=\protect\gamma_{M}=\protect%
\gamma_{R}=\protect\gamma=10^{-4}\protect\omega_{R}$, $T_L =0.2%
\protect\omega_{R}$, and $T_R =0.02\protect\omega_{R}$.}
\label{fig:amplifier}
\end{figure}

\textit{Amplifier}--The thermal amplifier means that a weak heat current at one terminal can
significantly \textit{amplify} the heat currents through the other two
terminals ones. The ability of the amplifier is quantified by the
amplification factor, for instance \cite{Joulain2016}, 
\begin{equation}
\alpha _{L,R}=\frac{\partial \dot{Q}_{L,R}}{\partial \dot{Q}_{M}},
\label{eq:amplificationfactor}
\end{equation}%
where  the heat current $\dot{Q}_M$ as the weak current to control
 the other two heat currents is implied. Strictly speaking, an amplifier is
achieved if the amplification factor $\alpha_{L,R}>1$. The larger the amplification factor is,
the better the amplification function is. We show the heat currents and the amplification
factors in Fig.~\ref{fig:amplifier}~(a) varying with $T_M$ for the internal
coupling strength $g=0.1\omega_{M}$ in the case of the steady state. It is obvious that 
$\dot{Q}_M$  is small in the reasonable range of the temperature $T_M$, while the other two
currents $\dot{Q}_L$ and $\dot{Q}_R$ are drastically changed. Fig.~\ref%
{fig:amplifier}~(b) shows that the amplification factor versus the temperature $T_M$. It can be easily found that
the absolute amplification factors $\left\vert\alpha_{L,R}\right\vert$ are about $10$ which shows that our 
system has the very strong amplification ability.

\begin{figure}[tbp]
\centering
\includegraphics[width=0.9\columnwidth]{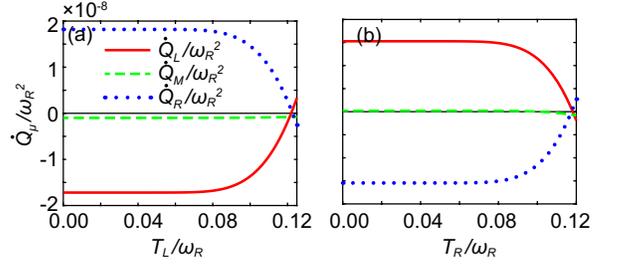}
\caption{(Colour online) Three heat currents $\dot{Q}_{\protect\mu}/\protect%
\omega_{R}^2$ (a) versus $T_{L}/\protect\omega_{R}$ and (b) versus $T_{R}/%
\protect\omega_{R}$ at the steady state. The red solid, green dashed, and blue
dotted lines correspond to the heat currents $\dot{Q}_{L}/\protect\omega_{R}^2$, $%
\dot{Q}_{M}/\protect\omega_{R}^2$, and $\dot{Q}_{R}/\protect\omega_{R}^2$,
respectively. Here $\protect\omega_{L}=0.9\protect\omega_{R}$, $\protect%
\omega_{M}=0.1\protect\omega_{R}$, $g=0.8\protect\omega_{M}$, $\protect\gamma%
_{L}=\protect\gamma_{M}=\protect\gamma_{R}=\protect\gamma=10^{-4}%
\protect\omega_{R}$. In addition, in (a) $T_R =0.12\protect\omega_{R}$, $T_{M}=0.08\protect%
\omega_{R}$, and in (b) $T_L =0.12\protect\omega_{R}$, $T_{M}=0.08\protect\omega%
_{R}$.}
\label{fig:stabilizer}
\end{figure}

\textit{Modulator}--A thermal modulator is used to \textit{modulate} the heat currents continuously
such that they can be changed from a small value to a large one by controlling a weak heat current. One can see from
Fig.~\ref{fig:amplifier} that with the slight change of the heat current $\dot{Q}_M$, the heat currents $\dot{Q}_{L,R}$ have been
 modulated from almost \textit{zero} value at a low temperature $T_M$ to a large
value at a high temperature $T_M$. In fact, such a phenomenon can also be found in Fig. \ref{fig:stabilizer} where the heat currents
  $\dot{Q}_{L,R}$ vary from a tiny value to a large value.  These are just the modulation function.

\textit{Switcher}-- The heat switcher can ``cut off" the heat currents if one weak heat current becomes weak enough. This function is quite obvious from Fig.~\ref{fig:amplifier}~(a) and  Fig. \ref{fig:stabilizer}. In Fig.~\ref{fig:amplifier}~(a), one can see that the currents will almost vanish when the temperature $T_M$ at the ``control" terminal become small. Similarly, in Fig. \ref{fig:stabilizer}, the heat currents will reach zero when the temperature $T_{L/R}$ as the ``control" terminal approaches a given value.

\textit{Stabilizer}--The feature of a heat stabilizer is that the heat currents are robust to the fluctuation of the temperature. Here we will show that our model can work as a stabilizer because the currents are not sensitive to the change of the temperature $T_L$ or $T_R$. As displayed
in Fig.~\ref{fig:stabilizer} (a), the three currents are obviously not sensitive to the change of the
temperature $T_L$ over the large range from $T_{L} =0$ to
around $T_{L} =0.8\omega_{R}$. The reason is that the greatly separated transition
frequency $\omega_{L}=0.9\omega_{R}$ is much larger than $T_{L}$ which prevents the
bath $L$ to drastically excite the qubit $L$'s transition, which can be
verified from Eqs.~(\ref{eq:JP})-(\ref{eq:photonnumber}). This situation is
also suitable for $T_R$ as presented in Fig.~\ref{fig:stabilizer} (b).
Although the coupling strength is set as $g=0.8\omega_{M}$, one can check that $stabilizer$
function still exists for relatively weak  coupling $g$ which shows that the contribution of
the interaction has the limited influence on the stabilizer compared with the difference of the transition frequencies.

\begin{figure}[tbp]
\centering
\includegraphics[width=0.9\columnwidth]{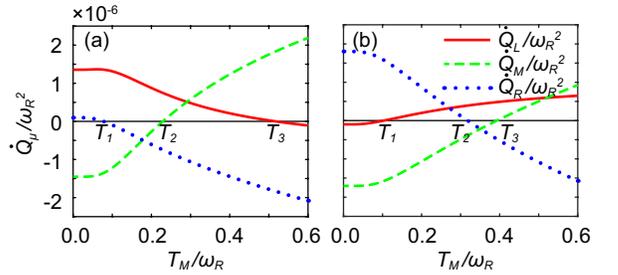}
\caption{(Colour online) Three heat currents $\dot{Q}_{\protect\mu}/\protect%
\omega_{R}^2$ versus $T_{M}/\protect\omega_{R}$ at the steady state. The red
solid, green dashed, and blue dotted lines correspond to currents $\dot{Q}%
_{L}/\protect\omega_{R}^2$, $\dot{Q}_{M}/\protect\omega_{R}^2$, and $\dot{Q}%
_{R}/\protect\omega_{R}^2$, respectively. Here we have $\protect\omega_{L}=0.6%
\protect\omega_{R}$, $\protect\omega_{M}=0.4\protect\omega_{R}$, $g=0.8%
\protect\omega_{M}$ fixed. In (a) the spontaneous decay rate $\gamma_{\mu}(\omega_{\mu l})=\gamma_{\mu}$ is chosen to be independent of frequency as mentioned before, i.e., $\protect\gamma_{L}=\protect\gamma_{M}=\protect\gamma_{R}=\protect\gamma=10^{-4}\protect\omega_{R}$, and $T_L =0.25\protect\omega_{R}$, $T_{R}=0.2\protect\omega_{R}$, while in (b) the Ohmic spectrum is applied \cite{Carmichael2002}, i.e., $\gamma_{\mu}(\omega_{\mu l})=\gamma_{\mu}\omega_{\mu l}$, and $\gamma_{\mu}$ is constant as above. The temperatures are set as $T_L =0.45\protect\omega_{R}$, $T_{R}=0.4\protect\omega_{R}$. Obviously the valve function is also obtained at  different critical temperature points. }
\label{fig:valve}
\end{figure}

\textit{Valve}--In analogy to a classical valve, a quantum thermal valve can
 perfectly cut off the
heat current at any one terminal and allow the heat to flow through the
other two terminals. Here we consider the heat current $\dot{Q}_M$ as the control terminal.
In order to see the valve function, we plot the three heat currents with respect to the temperature $%
T_M$ in Fig.~\ref{fig:valve}. As we see in subfigure (a) where the spontaneous decay rate is set independent of frequency, when
the temperature $T_M$ approaches to a critical temperature (about $0.09$),  the heat current $\dot{Q}_R$ is cut off and the 
heat can freely flow through the other two terminals. When $T_M$ reaches another critical temperature (about $0.24$), the heat current $\dot{Q}_M$ is cut off. When the $T_M$ reaches the third critical temperature (about $0.53$), the heat current $\dot{Q}_L$ is cut off.
In fact, the direction of the currents
can also be easily switched by controlling one temperature $T_M$. If the Ohmic spectrum of the bath is chosen, similar valve function can also be achieved as shown in subfigure (b) with different appropriate temperatures.

\begin{figure}[tbp]
\centering
\includegraphics[width=0.9\columnwidth]{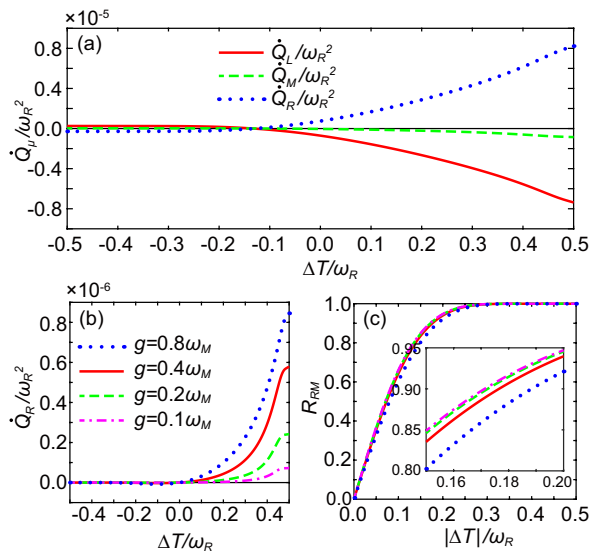}
\caption{(Colour online) (a) Three heat currents $\dot{Q}_{\mu}/\omega_{R}^2$ versus $\Delta T/\omega_{R}=(T_R - T_M)/\omega_{R}$ with $g=0.8\omega_{M}$ and $T_{L}=0.18\omega_{R}$. (b) Heat current $\dot{Q}_{R}/\omega_{R}^2$ versus $\Delta T/\omega_{R}=(T_R - T_M)/\omega_{R}$ for different coupling strength $g$ without the bath $L$. (c) The rectification factors $R_{RM}$ versus $|\Delta T |/ \omega_{R}=|(T_R - T_M)|/\omega_{R}$ corresponding to the cases in (b), and the inset is a close-up image over the range of $0.15<|\Delta T|/\omega_{R}<0.2$. In all the cases the spontaneous decay rates $\gamma_{L}=\gamma_{M}=\gamma_{R}=\gamma=10^{-4}\omega_{R}$, $\omega_{L}=0.9\omega_{R}$, $\omega_{M}=0.1\omega_{R}$, and the average temperature $T_A=(T_R+T_M)/2=0.25\omega_{R}$ are fixed. The blue dotted, red solid, green dashed, and magenta dot-dash lines correspond to $g=0.8\omega_{M}$, $g=0.4\omega_{M}$, $g=0.2\omega_{M}$, and $g=0.1\omega_{M}$ in (b) and (c), respectively.}
\label{fig:rectifier}
\end{figure}

\textit{Rectifier}--The significant feature of a rectifier is to allow the thermal
current flow unidirectionally, which is an analogue of the classical rectifier of the electricity. In Fig.~\ref{fig:rectifier} (a),
we plot the heat currents  at the terminals $R$ and $M$ versus the temperature difference $\Delta T= T_R -T_M$. It is obvious that when $\Delta T$ is larger than a critical value ($\sim-0.1$), the heat flows along a certain direction, for example, the heat flows out of the bath $R$ and  into the bath $M$ (as well as bath $L$). Here we'd better consider the terminal $M$ and $L$ as a whole to serve as the common terminal. On the contrary, if $\Delta T$ is less than the critical value, the almost vanishing heat will flow oppositely. So our system can work as a rectifier. However, one can find that the critical value does not lie at the \textit{zero} temperature difference, which means that swapping the heat baths $R$ and $M$ cannot change the direction of heat current with a small temperature difference. This is actually due to the existence of the third heat current $\dot{Q}_L$. To avoid this effect, we remove the bath $L$ so that the critical temperature difference can be translated to zero value. Meanwhile, our system can become a two-terminal quantum thermal device \cite{Maznev2013,Ordonez-Miranda2017}. In this case, the rectification factor, quantifying 
the ability of rectification, can be well defined as  \cite{Marcos-Vicioso2018} 
\begin{equation}
R=\frac{|\dot{Q}_{fore} -\dot{Q}_{back}|}{|\dot{Q}_{fore} +\dot{Q}_{back}|}.
\end{equation}%
The larger $R$ signifies the better rectification ability, and a perfect rectifier
is obtained for $R=1$.
 In Fig.~\ref{fig:rectifier} (b) we plot the heat current $\dot{Q}_R$ with respect to the temperature
difference $\Delta T $ for different $g$ and different $%
\omega_L$. It is obvious that the considerable heat current can only flow along a single direction, and the strong internal coupling is more beneficial to the large unidirectional heat current.
 In Fig.~\ref{fig:rectifier} (c), the rectification factors $R_{RM}$
 corresponding to Fig.~\ref{fig:rectifier} (b)  versus
the absolute temperature difference $|\Delta T|$ are plotted. Note that we have let $%
\dot{Q}_{fore}=\dot{Q}_{R}$ for $T_R > T_M$ while $\dot{Q}_{back}=-\dot{Q}%
_{R}$ for $T_R < T_M$.  %
We can also notice that large temperature difference results in almost
perfect rectification. Similarly the rectification effect can also be found
if the bath $R$ is removed.

\section{Discussion\label{sec:discussion}}

Before the end, we would like to give an intuitive but rough understanding to our device. A helpful way is to image our device has only three eigenfrequencies (levels) which satisfy the resonance condition, but two of which are extremely different.  It is natural that two eigenfrequencies  are relatively close to each other. Suppose each transition is driven by a thermal bath.  Such a configuration is much like the quantum refrigerator presented in Ref. \cite{Linden2010}. Due to the energy conservation, the output heat current released only by the transition subject to the maximal eigenfrequencies must be the same as the total input heat currents shared by the other two transitions subject to quite different eigenfrequencies. Since the two relatively close eigenfrequencies are separated in the input and the output terminals respectively, they should govern the similar large magnitude of the change of the heat currents. Correspondingly, the heat current at the third terminal will be only slightly changed. In the different parameter regimes, the input and output heat currents will be shared by different combinations of the three eigenfrequencies. So various interesting functions will appear. Compared with our current device with eight eigenfrequencies, 
it  includes many similar three-level transitions as above. However, they don't work separately (which directly leads to the difficulty to directly understand our device physically). Their cooperative effect can lead to much more complicated transitions and hence could enhance or reduce the working mechanism mentioned for the three-level case. In one word, the plentiful functions result from the asymmetry and the complexity (the strong internal coupling) of the covered transitions in the system, while one specific function made to be superior to the others results from the proper choice of the parameter regimes. 

Furthermore, one has to note that designing our quantum thermal device is greatly related to the choice of the system's structure and dissipation channels including the temperatures of the baths. As we know, both the transistor and the rectifier need the asymmetry of transition frequencies at the different terminals, the stabilizer requires the working transition frequencies are much larger than their corresponding  temperature in terms of numerical value, and the valve only needs that the heat currents can selectively vanish. Whether these functions above can be realized in some simpler systems such as a single qubit, qutrit or two qubits system remains an interesting question. However, one can easily check that, given the same spontaneous decay rates, the rectifier and the transistor are hard to implement in qubit systems \cite{Segal200501} due to lack of the asymmetric level configuration. The valve cannot be implemented in a qutrit due to the heat currents generally vanish simultaneously. In addition, the multi-level system of a single qudit (e.g. a qutrit) usually leads to the cross couplings between a single transition with different baths.  For the two-qubit system such as Ref.~\cite{Man2016},  the transistor function is not found (Here we do not use the rotating wave approximation), the reason is that two baths have to share the same dissipation channels via a single qubit, otherwise,  the cross coupling could also be covered. 

In addition, we want to emphasize that the choice of the three qubits' transition frequencies is related to the validity of global master equation and the system's thermal function. On one hand, the secular approximation has to be satisfied in the derivation of global master equation as shown in \cite{Seah2018,Gonzalez2017,Hofer2017NJP,Breuer2002}. It means the energy gap should be much greater compared with the system's decay rates. Any two qubits in our model possessing the same transition frequencies will generate the degenerate levels which result in the  violation of  the secular approximation no matter how strong the internal coupling is. On the other hand, there are only three energy configurations, i.e., $\omega_R$ equal to, larger or smaller than $\omega_L + \omega_M$ given three baths can be at any temperature. Similar functions can also be realized in these cases. What we should pay attention to is cautiously arranging their energy distribution especially for the transistor and rectifier that need greatly asymmetric energy levels as well known.

Finally, we will give a brief discussion about the possible experimental realization. As well known, a spin-like system can be easily realized, the key is how to realize trilinear interaction in a system. In Refs. \cite{Reiss2003,Pachos2004,Bermudez2009}, the authors have proposed how to construct a system with internal trilinear interaction. Especially in \cite{Bermudez2009}, Bermudez \textit{et al.} employed many spin-like trapped ions to construct a Hamiltonian including bilinear and trilinear coupling  by modifying external fields. The transition frequency of each ion and both the coupling strengths can be carefully adjusted. Zero strength of the bilinear coupling leads to our model. The coupling between the system and a bath can be achieved via a resonator, and a resistor act as a bath\cite{Karimi2017,Cottet7561}. In fact, the reservoir could be directly tailored with the desired bath spectrum by reservoir engineering, which has been described in detail and applied in many cases\cite{Scovil1959,Gelbwaser-Klimovsky2013,Myatt2000,Groblacher2015}. Note that as our model is general and the valid temperature regime is related to the choice of a qubit's frequency, so one can modify the desired temperature according to the qubit's frequency. 

\section{Conclusion\label{sec:conclusion}}
In this paper, a multifunctional quantum thermal device has been designed by utilizing   three resonantly and strongly  internal coupling qubits in contact with three heat baths. We study transport properties by applying the secular master equation. The steady-state thermal behaviours show that this thermal device can work as a thermal transistor, a switcher, a valve and even a thermal rectifier.
We would like to emphasize that the plenty of the functions mainly originate from the large difference between the transition frequencies of the qubits.  Whether it could induce some more novel applications is worthy of being studied in the future.

\section*{ACKNOWLEDGEMENTS}
This work was supported by the
National Natural Science Foundation of China, under Grant No.11775040 and
No. 11375036, and the Fundamental
Research Fund for the Central Universities under Grants No. DUT18LK45.

\appendix

\section{The eigen-decomposition of $H_S$ and the eigen-operators}

For the Hamiltonian $H_S$, the eigen-decomposition reads $%
H_S=\sum_k\lambda_k\left\vert
\lambda_k\right\rangle\left\langle\lambda_k\right\vert$, where the
eigenvalues $\lambda_k$ are given in the main text and the eigenvectors are
given as follows. {\allowdisplaybreaks[4] 
\begin{align}
\left\vert \lambda _{1}\right\rangle & =-\sin \theta _{1}\left\vert
111\right\rangle +\cos \theta _{1}\left\vert 000\right\rangle , \notag \\
\left\vert \lambda _{2}\right\rangle & =-\sin \theta _{2}\left\vert
101\right\rangle +\cos \theta _{2}\left\vert 010\right\rangle , \notag \\
\left\vert \lambda _{3}\right\rangle & =-\cos \theta _{3}\left\vert
100\right\rangle +\sin \theta _{3}\left\vert 011\right\rangle , \notag \\
\left\vert \lambda _{4}\right\rangle & =-\cos \theta _{4}\left\vert
110\right\rangle +\sin \theta _{4}\left\vert 001\right\rangle , \notag \\
\left\vert \lambda _{5}\right\rangle & =+\sin \theta _{4}\left\vert
110\right\rangle +\cos \theta _{4}\left\vert 001\right\rangle ,  \\
\left\vert \lambda _{6}\right\rangle & =+\sin \theta _{3}\left\vert
100\right\rangle +\cos \theta _{3}\left\vert 011\right\rangle , \notag \\
\left\vert \lambda _{7}\right\rangle & =+\cos \theta _{2}\left\vert
101\right\rangle +\sin \theta _{2}\left\vert 010\right\rangle , \notag \\
\left\vert \lambda _{8}\right\rangle & =+\cos \theta _{1}\left\vert
111\right\rangle +\sin \theta _{1}\left\vert 000\right\rangle , \notag 
\end{align}%
}%
with $\sin \theta _{i}=g/\sqrt{[\sqrt{(\mathit{\Lambda }_{i}^{2}+g^{2})}+%
\mathit{\Lambda }_{i}]^{2}+g^{2}}$, $\cos \theta _{i}=\sqrt{1-{\sin }%
^{2}\theta _{i}}$, ${\left\vert 1\right\rangle }_{\mu }={[1,0]}^{T}$ and ${%
\left\vert 0\right\rangle }_{\mu }={[0,1]}^{T}$ representing the excited and
the ground states of the $\mu $th qubit.
Thus the transition operators of the qubits can also be rewritten in the $H_S$ representation as
{\allowdisplaybreaks[4] 
\begin{align}
V_{L1}& = \sin \alpha _{L1}(\left\vert \lambda _{6}\right\rangle \left\langle
\lambda _{8}\right\vert -\left\vert \lambda _{1}\right\rangle \left\langle
\lambda _{3}\right\vert), \notag \\
V_{L2}& =\sin \alpha _{L2}(\left\vert \lambda _{1}\right\rangle
\left\langle \lambda _{6}\right\vert +\left\vert \lambda _{3}\right\rangle
\left\langle \lambda _{8}\right\vert ), \notag \\
V_{L3}& =\sin \alpha _{L3}(\left\vert \lambda _{5}\right\rangle \left\langle
\lambda _{7}\right\vert -\left\vert \lambda _{2}\right\rangle \left\langle
\lambda _{4}\right\vert  ), \notag \\
V_{L4}& =\sin \alpha _{L4}(\left\vert \lambda _{2}\right\rangle
\left\langle \lambda _{5}\right\vert +\left\vert \lambda _{4}\right\rangle
\left\langle \lambda _{7}\right\vert ), \notag \\
V_{M1}& =\sin \alpha _{M1}(\left\vert \lambda _{1}\right\rangle \left\langle
\lambda _{2}\right\vert +\left\vert \lambda _{7}\right\rangle \left\langle
\lambda _{8}\right\vert ), \notag \\
V_{M2}& =\sin \alpha _{M2}( \left\vert \lambda _{2}\right\rangle
\left\langle \lambda _{8}\right\vert -\left\vert \lambda _{1}\right\rangle
\left\langle \lambda _{7}\right\vert), \\
V_{M3}& =\sin \alpha _{M3}(\left\vert \lambda _{3}\right\rangle \left\langle
\lambda _{4}\right\vert +\left\vert \lambda _{5}\right\rangle \left\langle
\lambda _{6}\right\vert ), \notag  \\
V_{M4}& =\sin \alpha _{M4}(\left\vert \lambda _{3}\right\rangle
\left\langle \lambda _{5}\right\vert -\left\vert \lambda _{4}\right\rangle
\left\langle \lambda _{6}\right\vert ), \notag \\
V_{R1}& =\sin \alpha _{R1}(\left\vert \lambda _{5}\right\rangle \left\langle
\lambda _{8}\right\vert +\left\vert \lambda _{1}\right\rangle \left\langle
\lambda _{4}\right\vert ), \notag \\
V_{R2}& =\sin \alpha _{R2}(\left\vert \lambda _{1}\right\rangle \left\langle
\lambda _{5}\right\vert -\left\vert \lambda _{4}\right\rangle \left\langle
\lambda _{8}\right\vert ), \notag  \\
V_{R3}& =\sin \alpha _{R3}(\left\vert \lambda _{6}\right\rangle \left\langle
\lambda _{7}\right\vert +\left\vert \lambda _{2}\right\rangle \left\langle
\lambda _{3}\right\vert ), \notag \\
V_{R4}& =\sin \alpha _{R4}(\left\vert \lambda _{2}\right\rangle \left\langle
\lambda _{6}\right\vert-\left\vert \lambda _{3}\right\rangle \left\langle
\lambda _{7}\right\vert), \notag
\end{align}%
}%
where 
{\allowdisplaybreaks[4] 
\begin{align}
\alpha _{Lk} &=\frac{\mathrm{\pi }}{4}-\left( -1\right) ^{k}\left(\frac{\mathrm{\pi }}{4}-(\theta _{[k]}-\theta _{[k]+2})\right), \notag \\
\alpha _{Mk} &=\frac{\mathrm{\pi }}{4}-\left( -1\right) ^{k}\left(\frac{\mathrm{\pi }}{4}-(\theta _{[k]}-\theta _{[k]+1})\right),  \\
\alpha _{Rk} &=\frac{\mathrm{\pi }}{4}+\left( -1\right) ^{k}\left(\frac{\mathrm{\pi }}{4}-(\theta _{[k]}+\theta _{5-[k]})\right), \notag
\end{align}%
}%
with $[k]$ denoting the minimal integer not less than $k/2$,
and the corresponding eigenfrequencies are given by

{\allowdisplaybreaks[4] 
\begin{align}
\omega _{L1,2}&=\sqrt{\mathit{\Lambda }_{1}^{2}+g^{2}}\mp \sqrt{\mathit{%
\Lambda }_{3}^{2}+g^{2}}, \notag \\
\omega _{L3,4}&=\sqrt{\mathit{\Lambda }_{2}^{2}+g^{2}}\mp \sqrt{\mathit{%
\Lambda }_{4}^{2}+g^{2}}, \notag \\
\omega _{M1,2}&=\sqrt{\mathit{\Lambda }_{1}^{2}+g^{2}}\mp \sqrt{\mathit{%
\Lambda }_{2}^{2}+g^{2}}, \notag \\
\omega _{M3,4}&=\sqrt{\mathit{\Lambda }_{3}^{2}+g^{2}}\mp \sqrt{\mathit{%
\Lambda }_{4}^{2}+g^{2}}, \\
\omega _{R1,2}&=\sqrt{\mathit{\Lambda }_{1}^{2}+g^{2}}\mp \sqrt{\mathit{%
\Lambda }_{4}^{2}+g^{2}}, \notag \\
\omega _{R3,4}&=\sqrt{\mathit{\Lambda }_{2}^{2}+g^{2}}\mp \sqrt{\mathit{%
\Lambda }_{3}^{2}+g^{2}}. \notag
\end{align}
}%
 It is obvious that
the eigenoperators and their corresponding eigenfrequencies $\omega _{\mu l}$
satisfy $[H_{S},V_{\mu l}(\omega _{\mu l})]=-\omega _{\mu
l}V_{\mu l}(\omega _{\mu l})$.
\bibliography{Three-qubits-model}
% Produces the bibliography via BibTeX.
%\onecolumngrid

\end{document}